\documentclass[12pt]{iopart}

\usepackage{graphicx}
\usepackage{epsfig}
\usepackage{txfonts}

\begin{document}

\title{Fundamental constraints on particle tracking with optical tweezers}  
\author{Michael~A~Taylor, Joachim~Knittel and Warwick~P~Bowen}

\address{Centre for Engineered Quantum Systems, University of Queensland, St Lucia, Queensland 4072, Australia}
\date{\today}

\begin{abstract}
 A general  quantum limit to the sensitivity of particle position measurements is derived following the simple principle of the Heisenberg microscope. The value of this limit is  calculated for particles in the Rayleigh and Mie scattering regimes, and with parameters which are relevant to optical tweezers experiments. The minimum power required to observe the zero-point motion of a levitating bead is also calculated, with the optimal particle diameter always smaller than the wavelength. We show that recent optical tweezers experiments are within two orders of magnitude of quantum limited sensitivity, suggesting that quantum optical resources may soon play an important role in high sensitivity tracking applications.
\end{abstract}


\maketitle       

\section{Introduction}          


 In optical tweezers, the electric field gradient of a tightly focused laser beam is used to trap small particles~\cite{Ashkin1987}. In this interaction, momentum is imparted to the trapped object when it scatters photons from the incident optical field. The particle position can then be determined by measuring these scattered photons at the back-focal plane of a condenser lens, typically with a quadrant detector~\cite{Gittes1998}. This simple concept allows particles to be tracked with sub-nanometer sensitivity~\cite{Chavez2008} as forces ranging from subpiconewton~\cite{Moffitt2008,Finer1994} to nanonewton~\cite{Jannasch2012} are controllably applied. Applications of this technique range over many fields, such as the study of Brownian motion~\cite{Franosch2011,Huang2011} and optomechanics~\cite{Chang2010,Li2011}, but it has been most significant to subcellular biology. The biophysics which has been revealed through optical tweezers includes both the dynamics and magnitude of the forces applied by biological motors~\cite{Finer1994,Svoboda1993,Bustamante2004},  the stretching and folding properties of DNA and RNA~\cite{Bustamante2004,Bustamante2005,Greenleaf2006}, the dynamics of virus-host coupling~\cite{Kukura2009}, the strain on an enzyme during catalysis~\cite{Bustamante2004}, and the rheological properties of cellular cytoplasm~\cite{Taylor2012,Yamada2000,Senning2010,Norrelykke2004}.

 Attaining high sensitivity is vital when exploring processes which lead to smaller and faster movements. The position sensitivity is usually limited by laser noise, electronic noise in the detector, or drifts of mirrors in the experiment. Substantial efforts have been made to minimize each of these sources of error~\cite{Chavez2008,Moffitt2008,Kukura2009,Taylor2011}. With sufficient improvements, the sensitivity per photon must eventually be limited by noise due to the quantization of light~\cite{Kolobov2000}. At this point, only two options remain to improve the sensitivity; either using more photons, which can damage biological samples~\cite{Peterman2003,Neuman1999}, or using photons with quantum correlations which allow the fundamental limit to be broken~\cite{Giovannetti2004,Brida2010,Treps2003}, as has recently been demonstrated in optical tweezers~\cite{Taylor2012}.

 The quantum noise limit is an important consequence of quantum mechanics, and is becoming increasingly relevant in experiments. So, what is the precision at the quantum limit? And how closely are experiments approaching this?  Until now, the quantum limit has only been derived numerically in the paraxial optics regime with Rayleigh scattering~\cite{Tay2009}, making the calculated limit inapplicable to most experiments and difficult to understand intuitively. Here we follow the simple principles of the Heisenberg Microscope to derive a quantum sensitivity limit in optical tweezers. This technique  can be straightforwardly applied to particles with arbitrary shape and size, yields analytic solutions in the Rayleigh scattering regime, and allows analysis of a wide range of optical setups.  The approach is also relevant to other measurements such as microscopy with fluorescent particles. Our results show that leading experiments are already within two orders of magnitude of the quantum noise limit, which suggests that future improvements may soon need to integrate quantum resources.

 

\section{Principle}

Particles in optical tweezers are tracked by measuring their perturbing influence on the electric field as they introduce scattered photons. The position and momentum of these scattered photons is indeterminate, as described by the infamous Heisenberg uncertainty relation~\cite{Giovannetti2004}
\begin{equation}
\Delta q_i \Delta p_i \geq \frac{\hbar}{2},\label{Heisenberg}
\end{equation}
where $\hbar$ is the reduced Planck constant, and $\Delta q_i $ and $\Delta p_i$ are respectively the standard deviations of the photon momentum and position at the scattering origin. This is defined along the axes $i\in \{ x,y,z\}$, where $z$ is the direction of optical propagation, and $x$ the axis of linear polarization.
We can use Eq.~\ref{Heisenberg} to place a bound on  $\Delta q_i$ in terms of the standard deviation of photon momentum $\Delta p_i$, which can be easily calculated. The  magnitude of this momentum deviation can be at most the total photon momentum $p_{ph} = \frac{2 \pi \hbar}{\lambda}$, where  $\lambda$ is the vacuum wavelength, because 
\begin{equation}
\Delta p^2_i =\langle p_i^2 \rangle-\langle p_i \rangle^2 \leq \sum_{i=x,y,z} \langle p_i^2 \rangle-\langle p_i \rangle^2= p_{ph}^2 - \sum_{i=x,y,z}\langle p_i \rangle^2 . \label{p_ph}
\end{equation}
 We introduce a parameter $f$ to characterize $\Delta p$ along each axis, such that
\begin{equation}
\Delta p_i =  f_i p_{ph} ,
\end{equation}
leading to the condition $f_x^2 +f_y^2+ f_z^2 \leq 1$. This equality is met for a photon which is scattered with no preferred direction $\left( \langle p_i \rangle=0 \right)$.  From Eq.~\ref{Heisenberg}, a measurement of the origin of a single scattered photon is constrained to
\begin{equation}
\Delta q_i \geq \frac{\lambda}{4 \pi f_i}.
\end{equation}
In particle tracking experiments, a measurement is typically performed with $N_s=\kappa N_0$ independent scattered photons, where $N_0$ incident photons have a scattering probability of $\kappa$. Since the photons are uncorrelated, the standard error in finding the scattering origin $\delta q_i$ is given by
\begin{equation}
\delta q_i \geq \frac{\lambda}{4 \pi f_i \sqrt{\kappa N_0}}. \label{QNL_eq}
\end{equation}
 This determines the accuracy with which the origin of a collection of scattered photon can be determined.  When the scattered photons originate at the center of the trapped particle, the particle can be tracked with a  position uncertainty given by Eq.~\ref{QNL_eq}. This is the case for Rayleigh scattering, where the scattered field is perfectly centered on the particle. In general, however, the mean origin of scattered photons can deviate slightly from the particle center, so Eq.~\ref{QNL_eq} defines a lower bound on particle position uncertainty. 

To derive the value of the quantum sensitivity limit given in Eq.~\ref{QNL_eq}, we only need to calculate the scattering probability $\kappa$ and the momentum parameter $f$. As these parameters depend on the trapping field, the particle being tracked, and the refractive index of the surrounding medium, they must be evaluated for specific cases,  although neither may ever reach unity. A general lower bound of $\delta q \geq \frac{\lambda}{4\pi\sqrt{N_0}}$ must always be imposed, even if the scattering properties of the particle were cleverly engineered~\cite{Jannasch2012} and the measurement performed perfectly; for all realistic situations, the minimum resolvable displacement is larger than this. To determine the limit for real experiments, we evaluate the quantum limit for the common case of homogeneous spheres in a Gaussian trap. 



\begin{figure}
 \begin{center}
   \includegraphics[width=8cm]{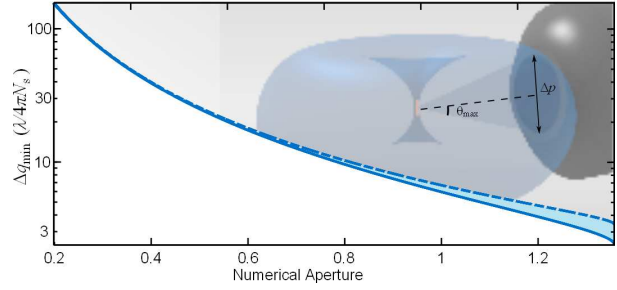}
   \caption{ Light scatters from a small particle with a defined spatial profile, shown here as Rayleigh scattering. The field which enters the objective aperture  is measured to determine the particle position. The scattered photons which propagate towards the aperture ($\theta > \theta_{\rm min}$) have a momentum distribution $\Delta p$ which defines the fundamental lower bound on sensitivity with which the particle position can be determined, as plotted here as a function of the condenser NA for Rayleigh particles suspended in water. Sensitivity improves with NA, as high NA lenses collect scattered photons with a larger momentum range.  Together with the scattering rate $\kappa$, this defines the quantum limit on displacement resolution per incident photon.
}
 \label{Rayleigh}  
 \end{center}
\end{figure}

\section{Rayleigh Scattering}

Both the scattering rate $\kappa$ and momentum distribution parameter $f$ can be calculated analytically in the Rayleigh scattering regime, in which the particle diameter $d \ll \lambda$. If an objective with numerical aperture of NA focuses an incident TEM$_{00}$ mode on the particle, the scattering rate is
%
\begin{equation}
\kappa = \frac{2}{3} \pi^5 {\rm NA}^2 \left(\frac{d}{\lambda}\right)^6 \left(\frac{ n_p^2 - n_m^2}{n_p^2 + 2 n_m^2} \right)^2,
\end{equation}
where $n_p$ and $n_m$ are respectively the refractive index of the particle and surrounding medium~\cite{Bohren}. For Rayleigh scatterers polarized along the $x$ axis, the scattered photons enter the mode $\psi$, with amplitude given by
\begin{equation}
|\psi|^2 = \frac{3}{8 \pi} \frac{1-(x/r)^2}{r^2},
\end{equation}
 where $r$ is the radial distance~\cite{Bohren}. Assuming the photons to be moving radially outward, this mode determines the momentum profile of the scattered light. In most experiments, any scattered photon which reaches the detector has passed through the aperture of an objective. This constrains the momentum range of the detected photons, thereby increasing the position uncertainty. The measurement then only includes the light present in the area $A$ of this aperture. For this measurement, the parameter $f$ is calculated to be
\begin{equation}
f_x^2 = \int |\psi|^2 x^2 dA -  \left( \int |\psi|^2 x dA \right) ^2 \label{d}.
\end{equation}
The second term in Eq.~\ref{d} is zero for all cases considered here, as photons scattered from centered particles have no preferred transverse direction. The ultimate sensitivity limit is given when we evaluate this over a spherical shell; here this gives $f_x^2=\frac{1}{5}$, $f_y^2=\frac{2}{5}$ and $f_z^2=\frac{2}{5}$. Since the sensitivity predicted from this can only be achieved with perfect imaging of all scattered photons, it also determines the measurement back-action upon the particle. For measurements subject to a limited aperture size,  Eq.~\ref{d} can be integrated analytically, with the corresponding quantum limit plotted in Fig.~\ref{Rayleigh}. In this, the aperture size is expressed in terms of the NA for a particle suspended in water. This shows near-perfect agreement with the equivalent numerical calculations performed in Ref.~\cite{Tay2009}. 



\section{Mie Scattering}

\begin{figure}
 \begin{center}
   \includegraphics[width=8cm]{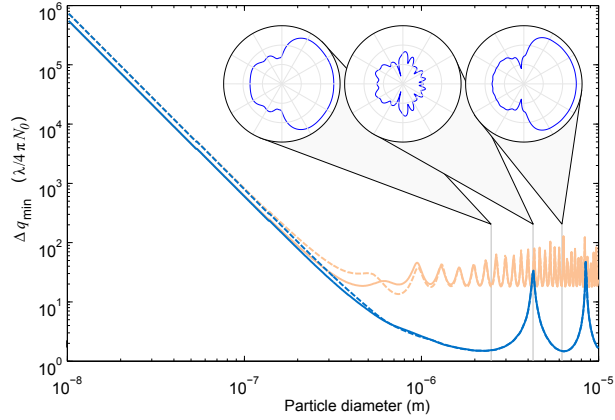}
   \caption{ The quantum limit on position sensing polystyrene beads ($n=1.58$) suspended in water with a 1.3 NA objective as a function of diameter. The faint and dark lines  representing measurement of back-scatter and forward-scatter respectively, and dashed and solid lines representing measurements parallel and perpendicular to the plane of polarization. 
Insets: Logarithmically scaled scattered field intensity profiles for bead diameters of 2.5, 4.3 and 6.2 $\mu$m, calculated in the plane of polarization. The 4.3 $\mu$m bead forward-scatters far less light than either of the other two, because resonant effects reduce the interaction of the incoming field with the bead, such that most of the light remains in the incident mode. However, the back-scatter is of a similar magnitude for all three.
}
 \label{Mie_QNL}  
 \end{center}
\end{figure}

 Optical tweezers experiments usually operate with particles which are too large to be accurately approximated as Rayleigh scatterers. For these particles, the scattering profile is complicated by such effects as multiple internal reflections and the interference between optical paths of different length. For spherical particles, this scattering regime is described mathematically by Mie theory~\cite{Bohren}. 
 We evaluate the scattering profiles numerically with the Optical Tweezers Toolbox\cite{Nieminen2007}. These profiles are integrated as described in Eq.~\ref{d} to find the quantum sensitivity limit for particle tracking. The quantum limit is shown in Fig.~\ref{Mie_QNL}  for polystyrene beads suspended in water, measured with 1064~nm light and an NA=1.3 objective and condenser. Examining this, we see a number of noteworthy features. For small particles, where Rayleigh scattering is valid, forward and back scatter offer an equally sensitive measurement. However, as the particle size increases, the scatter becomes preferentially forward, with less information carried by back-scattered photons. Polarization has little effect on the sensitivity for beads larger than the wavelength. 
 Two different resonant effects are evident in Fig.~\ref{Mie_QNL} for large particles.
The direction of scattering is periodically modulated by an effect which is approximately given by thin-film interference. This causes a significant modulation of the sensitivity available through back-scatter measurements; however, maximizing back-scattered sensitivity may not be advantageous for high-refractive index particles, as this regime also has the weakest trapping potential~\cite{Stilgoe2009}. In addition to this,  Mie resonances suppress the particle position information encoded in the phase of the outgoing field when propagation through the particle increases the optical path length by an integer multiple of the wavelength (see Fig.~\ref{Mie_QNL} insets). This degrades the measurement sensitivity without affecting the trapping potential, as the trapping is determined purely by the optical amplitude.

\section{Applications}
\subsection{Resolving zero-point motion}

\begin{figure}
 \begin{center}
   \includegraphics[width=8cm]{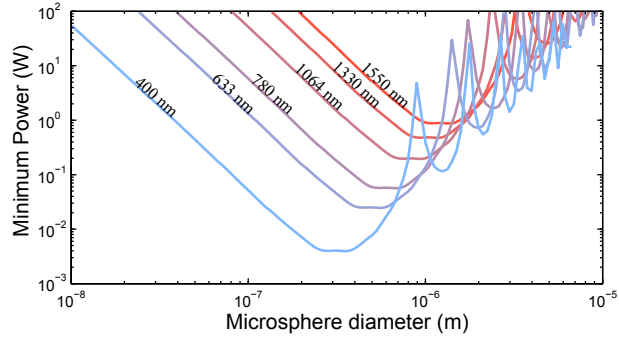}
   \caption{ 
This is the optical power required to observe the zero-point motion of a levitating silica bead in vacuum, for six different laser wavelengths. A refractive index of 1.46, a temperature of $T=298$~K, and a decay rate of $\Gamma=2 \pi \times 0.1$~Hz have been used, along with NA=0.8 objective and condenser. This plot can be easily extrapolated to other physical parameters because the required power scales linearly with both the decay $\Gamma$ and temperature $T$. Although the decay rate used for this calculation is similar to that in recent experiments~\cite{Li2011}, future experiments may attain much lower decay rates, with predicted rates of order $10^{-6}$~Hz~\cite{Chang2010,Geraci2010}.
}
 \label{ZPMplot}  
 \end{center}
\end{figure}

One application of this calculation is to determine the conditions required to observe the zero-point motion of a levitating sphere. When this can be measured, it is possible to generate squeezed mechanical states or cool to the ground state~\cite{Chang2010}, and also provides a means to search for the non-Newtonian gravity predicted at small scales~\cite{Geraci2010}. The average amplitude of zero-point motion is given by~\cite{Chang2010}
\begin{equation}
\Delta q_{\rm zpm} = \left( \frac{\hbar}{ 2 m \Omega} \right)^{1/2} \label{ZPM_eq}
\end{equation}
for a mechanical frequency $\Omega$ and mass $m$. For most interesting applications, the measurement time must be short compared to the average time for one bath phonon to enter or leave the mechanical mode~\cite{Chang2010}, which is given by $\tau=\frac{2 \pi}{\Gamma n}$ for a phonon occupancy $n=\frac{k_B T}{\hbar \Omega}$ and decay rate $\Gamma$. Although feedback is often used to cool motion, it does not influence this coupling rate, as feedback can only reduce the phonon occupancy by introducing  mechanical dissipation which increases $\Gamma$, keeping the product unchanged. In order to measure the movement $\Delta q_{\rm zpm}$ over a time $\tau$, the incident power must exceed the minimum threshold which we calculate here.  Conveniently, an increase in the mechanical frequency $\Omega$ reduces both the mechanical amplitude $\Delta q_{\rm zpm}$ and phonon occupancy $n$ such that the power threshold is independent of the  frequency. This means that the trapping potential need not be considered, and the derived power limit is valid even if there are additional trapping fields. The power threshold is calculated as a function of sphere radius and shown in Fig.~\ref{ZPMplot}. This shows that for a given laser, the optimal bead diameter is somewhat smaller than the wavelength. In the Rayleigh scattering regime, the required power scales as $\lambda^7$, making a short wavelength laser a practical choice. For large beads, the preferred wavelength is determined entirely by the Mie resonances. The dependence of the zero-point motion on particle mass generally makes its observation more difficult in larger beads.  To observe zero-point motion in any silica bead with a diameter larger than 6~$\mu$m would be prohibitively difficult at the assumed decay rate of 0.1~Hz, as it requires a quantum limited measurement of over 10~W of optical power.

Although the power predicted in Fig.~\ref{ZPMplot} should make the zero-point motion observable, it will also impart back-action onto the bead motion. The back-action will be greater than the zero-point momentum, since only the forward-scattered photons are included in the measurement. If the measurement contains any non-optimality, this will only enlarge the relative back-action. In real experiments, optical loss and the use of split detectors~\cite{Tay2009} ensure a non-optimal position measurement. This is a significant limitation, as measurement non-optimality prevents ground state cooling and minimum uncertainty squeezing~\cite{Naik2006}.

\subsection{Experimental sensitivity}


 With our calculations, we can characterize the gap between experimental results and the quantum limit. 
 %
%
 A recent experiment achieved a sensitivity of $1.7\times10^{-14}$~m/Hz$^{-1/2}$ when tracking 1~$\mu$m diameter polystyrene beads in water with 700~mW of 1064~nm light, and collecting only the forward-scatter~\cite{Chavez2008}. 
 Assuming the objective and condenser have NA=1.3, we find that $\kappa^{1/2} f_x=0.380$, and the $3.7\times 10^{18}$ incident photons per second should limit the sensitivity to $1.1\times 10^{-16}$~m/Hz$^{-1/2}$, which is within a factor of 145 of the demonstrated result.
In this experiment, the detected field was attenuated by $70\%$ before the detector, and the remaining photons were measured with sensitivity within a factor of 80 of the quantum limit.
 In another recent experiment, silica beads in vacuum were tracked at $3.9\times10^{-14}$~m/Hz$^{-1/2}$ using 120~mW of 1064~nm light, and NA=0.68 objective and condenser~\cite{Li2011}. For this case, $\kappa^{1/2} f_x=0.203$ and the quantum limit is $5.2\times 10^{-16}$~m/Hz$^{-1/2}$, only 75 times lower than the experimental sensitivity. In this experiment, the measurement for each axis was performed on separate detectors, so each detector had fewer photons available for measurement. 
  Although both of these experiments used non-optimal split detectors, they operate well within two orders of magnitude of the quantum limit. If further improvements to sensitivity are required, it would be useful to characterize the contributions to this gap arising from optical loss, detector non-optimality, and various noise sources. This characterization would indicate the aspects of the measurement which could be most effectively improved. Classical noise can be suppressed either by eliminating the noise or by using a stroboscopic measurement~\cite{Taylor2012}; otherwise, spatial homodyne~\cite{Tay2009} or self-homodyne detection~\cite{Taylor2012} can offer an optimal detection scheme, with quantum correlated light allowing still further improvement~\cite{Taylor2012}.

%



 The quantum limit calculated here has important implications for future experiments requiring better sensitivity than currently available. 
The thermal motion of particles in water has been observed to deviate from Brownian motion on very short time-scales, due to hydrodynamic resonances~\cite{Franosch2011} and ballistic motion between collisions~\cite{Huang2011}. However, there are further predicted effects which remain unobserved, such as oscillations arising from the elastic compressibility of water~\cite{Huang2011}. Observation of this exotic property of water requires sensitivity of around $2\times 10^{-17}$ m/Hz$^{-1/2}$~\cite{Huang2011}, which even with a quantum limited measurement requires 19~W of 1064~nm light. This presents a problem, as the surrounding water will boil when the optical power is around 10~W~\cite{Peterman2003}.  The only way to surpass the quantum limit is to use correlated photons which allow more information to be extracted per photon~\cite{Taylor2012,Kolobov2000,Giovannetti2004}. Such quantum resources can be integrated into particle tracking experiments~\cite{Taylor2012}, and with existing technology,  offer up to 10~dB of enhancement~\cite{LIGO2011}.  With this quantum enhancement, measurement of the elastic compressibility of water would require a more achievable 2~W of optical power. Based on our calculations, and the advanced state of recent experiments, we conclude that  quantum resources will play an important role in the next generation of high precision tracking experiments.



\section*{Acknowledgments}
We would like to thank Alexander Stilgoe and Timo Nieminen for advice about the Optical Tweezers Toolbox. This work was supported by the Australian Research Council Discovery Project Contract No. DP0985078.








\end{document}